\documentclass{INTERSPEECH2023}

\interspeechcameraready

\title{A Multi-dimensional Deep Structured State Space Approach to \\Speech Enhancement Using Small-footprint Models}

\name{Pin-Jui Ku$^1$,  Chao-Han Huck Yang$^1$, Sabato Marco Siniscalchi$^{1,2,3}$, Chin-Hui Lee$^1$}
\address{
  $^1$Georgia Institute of Technology, U.S.A\\\
  $^2$Kore University of Enna, Italy \\
  $^3$Norwegian University of Science and Technology, Norway}
\email{\{pku9,chl\}@gatech.edu}

\begin{document}
\maketitle

\begin{abstract}
We propose a multi-dimensional structured state space (S4) approach to speech enhancement. To better capture the spectral dependencies across the frequency axis, we focus on modifying the multi-dimensional S4 layer with whitening transformation to build new small-footprint models that also achieve good performance.
We explore several S4-based deep architectures in time (T) and time-frequency (TF) domains. The 2-D S4 layer can be considered a particular convolutional layer with an infinite receptive field although it utilizes fewer parameters than a conventional convolutional layer. Evaluated on the VoiceBank-DEMAND data set, when compared with the conventional U-net model based on convolutional layers, the proposed TF-domain S4-based model is 78.6\% smaller in size, yet it still achieves competitive results with a PESQ score of 3.15 with data augmentation. By increasing the model size, we can even reach a PESQ score of 3.18.
\end{abstract}


\section{Introduction}
\label{Sec:Intro}

%Speech enhancement aims at reducing noise in order to improve the quality and intelligibility of noisy speech, %.  This signal pre-processing technology also plays a critical role in building robust spoken language systems, such as speech recognition~\cite{Wu2017} and speaker recognition~\cite{Hansen2015} in adverse environments. %Before the rising of Deep Neural Network (DNN) based approaches in the past decade,
%it also plays a critical role in building robust speech technology, such as speech recognition~\cite{Wu2017} and speaker recognition~\cite{Hansen2015} in adverse environments.
%Conventional speech enhancement usually relies on linear statistical models to predict speech and noise. For instance, Spectral subtraction ~\cite{Boll1979} subtracts an estimate of the short-term noise spectrum to produce an estimated spectrum of clean speech. Wiener filtering ~\cite{Lim1979} estimates the clean speech by %statistically minimizing the mean-squared error between the original and enhanced speech. %Another well-known approach is
 %Kalman filtering ~\cite{Paliwal1987, Krishnan2006}  recursively estimates clean speech from noisy measurements, by leveraging upon the temporal evolution properties in speech.

Speech enhancement aims at reducing noise to improve the quality and intelligibility of noisy speech. It also serves as a front-end in robust speech recognition~\cite{Wu2017} and speaker recognition~\cite{Hansen2015} in adverse environments. Conventional approaches to speech enhancement usually rely on using statistical models to predict speech and noise, e.g., spectral subtraction ~\cite{Boll1979}, Wiener filtering ~\cite{Lim1979}, and Kalman filtering ~\cite{Paliwal1987, Krishnan2006}. In recent years, deep neural network (DNN) based approaches have demonstrated superiority over conventional signal processing techniques, especially for alleviating problems caused by musical and non-stationary noise ~\cite{Berouti1979}. They can be roughly categorized according to the features used. In the time-frequency (TF) domain, two distinct types exist: (i) spectral mapping ~\cite{Xu2014, Xu2015}, by establishing a regression function directly from mapping the noisy speech spectrogram to the clean one;  and (ii) masking~\cite{wang2008time, luo2019conv} by estimating a 2-D ratio mask that describes the TF relationships between clean speech and background noise. More recently, %researchers have proposed methods based  on
generative models, aiming at learning a prior distribution over clean speech data,  have been proposed, e.g., generative adversarial networks (GAN)~\cite{Pascual2017, Fu2019, Fu2021}, variational autoencoders (VAE)~\cite{Bando2020, Fang2021}, and diffusion models~\cite{Lu2022, richter2022, yen2022}.

DNNs~\cite{Xu2015, Grais2014, Wang2014}, recurrent neural networks (RNNs)~\cite{Gao2018, Huang2014, Weninger2015}, and convolutional neural networks (CNNs)~\cite{Giri2019, Dfossez2020} have been proposed for speech enhancement.  While RNNs are a natural choice for sequential inputs, their recursive nature makes them slow to train and suffer from optimization difficulties, such as the ``vanishing gradient'' problem~\cite{Pascanu2013}, limiting their ability to handle long sequences. In contrast, CNNs encode local context and enable parallel training, but they are fundamentally constrained by the size of their receptive field and may not achieve global coherence. Recently, deep state space models (SSM)~\cite{gu2021combining}, in particular S4-based ones~\cite{Gu2022efficiently}, have achieved state-of-the-art results in modeling sequence data with extremely long-range dependencies. The S4 layer unifies the strengths of both RNN and CNN layers: (i) it allows parallel computing like CNN layers, and (ii) it can capture global information across the whole input sequence like RNNs. S4-based systems have been deployed for raw audio generation~\cite{Gu2022Sashimi} and speech recognition~\cite{miyazaki2022}.% and we here propose to use it for speech enhancement. %as well as the historical relationship between SSM and SE, we conducted a series of investigations to determine how to apply S4 to the SE task.

%Although the S4-based model proposed in~\cite{Gu2022Sashimi} can be used as a time-domain regression model to achieve good baseline results,  it has been observed by \cite{So2011} that the Kalman filter with low order linear predictor is more suitable for enhancing slow-changing signals such as short-time Fourier transform (STFT) features than for enhancing the speech signal in the time domain directly. For this reason, we have shifted our focus from the time  domain to the TF domain with the hope of further improvement. However, the original S4-based model is designed to handle only 1D signals, and directly feeding the 2-D spectrograms into the model results in unsatisfactory performance, as observed in our experiments, since the model couldn't capture the harmonic dependencies across the frequency axis. To tackle this problem, we first compute the covariance matrix of each frequency bin and apply the whitening technique in order to make every frequency bin statistically independent. Unfortunately, While whitening is effective under certain scenarios, the overall performance remains non-ideal. Our next objective is to leverage the multi-dimensional S4 (S4ND)~\cite{Nguyen2022snd}, which expands the modeling capacity of S4 to multi-dimensional data, to create a new SE model. Following several unsuccessful attempts, we have successfully introduced a new U-net-like SE model that achieves better PESQ scores than the original time-domain model while having significantly fewer parameters (only 0.75M).

In this study, we propose a structured state-space-based architecture in the multi-dimensional TF domain for speech enhancement. We explore 2-D extensions since the original S4-based layer was designed to handle only 1-D input signals. However, to adequately capture the spectral dependencies across the frequency axis, we first compute the covariance matrix of each frequency bin and apply a whitening technique to make the frequency bins statistically uncorrelated. Unfortunately, while whitening was effective under specific scenarios, the overall performance improvement was limited. We therefore utilize the multi-dimensional S4 layer proposed in ~\cite{Nguyen2022snd} and adapt it to the TF domain. Experimental evidence shows that by injecting the multi-dimensional S4 layer inside a U-net-like architecture, we were able to deploy a novel deep enhancement model that attained better PESQ scores than the original time-domain model while reducing the model size by 78.6\% (with only 0.75M parameters). Since S4 layer directly handles 1-D inputs, we have also built a sequence-to-sequence regression model in the time-domain. Furthermore, we have also evaluated the effect of data augmentation on speech enhancement.

In the following sections, we will demonstrate three main contributions. We first conduct an initial investigation on how to best incorporate an S4 layer into speech enhancement models. To the best of our knowledge, this is the first attempt in this field. Next, we find that  TF-domain models can be improved using a small hop-length to extract spectrograms with long time lengths through short-time Fourier transform (STFT) mainly because S4 can easily handle long time-series inputs. Finally, by unifying the strengths of RNN and CNN layers, we achieve competitive enhancement results even with a compact TF-domain S4-based U-Net architecture

\section{Related Work}

\subsection{SSM - State Space Modeling}
\label{subsec:1DSSM_intro}
The S4 layer is based on the linear state space layer (LSSL) proposed in ~\cite{gu2021combining}, where  $u(t) \in \mathcal{C}$, and $v(t) \in \mathcal{C}$ are 1-dimensional input/output patterns in state space modeling, respectively, and $x(t) \in \mathcal{C}^N$ is an implicit \textit{N}-dimensional state vector. LSSL defines a 1-D mapping function $u(t) \rightarrow v(t)$ leveraging upon the following ordinary differential equations (ODEs):
\vspace{-1.0mm}
\begin{equation} \label{eq:continuous-SSM}
    \begin{aligned}
        \dot{x}(t) = \boldsymbol{A} x(t) +\boldsymbol{B} u(t) \\
        v(t) = \boldsymbol{C} x(t) + \boldsymbol{D} u(t)
    \end{aligned}
\end{equation}
Eq. (\ref{eq:continuous-SSM}) describes a state space model (SSM). The goal of LSSL is to treat the SSM as a black-box representation in a deep model, where the state-transition matrix $\boldsymbol{A} \in \mathcal{C}^{N\times N}$ and the projection matrices $\boldsymbol{B} \in \mathcal{C}^{1 \times N}, \boldsymbol{C} \in \mathcal{C}^{N \times 1}, \boldsymbol{D} \in \mathcal{C}^{1 \times 1}$ are parameters that could be learned by gradient descent.
To use LSSL in a discrete sequence-to-sequence model, the bilinear method~\cite{Tustin1947} can be used to discretize all continuous-time signals as follows:
\vspace{-1.0mm}
\begin{equation} \label{eq:discrete-SSM}
\begin{aligned}
    &x_{k} = \boldsymbol{\overline{A}}x_{k-1} + \boldsymbol{\overline{B}}u_k \\
    &v_k = \boldsymbol{C}x_{k} + \boldsymbol{D}u_k \\
    &\boldsymbol{\overline{A}} = (\boldsymbol{I} - \Delta / 2 \cdot \boldsymbol{A})^{-1}(\boldsymbol{I} + \Delta / 2 \cdot \boldsymbol{A})\\
    &\boldsymbol{\overline{B}} = (\boldsymbol{I} - \Delta / 2 \cdot \boldsymbol{A})^{-1}\Delta\boldsymbol{B}
\end{aligned}
\end{equation}
where $\Delta$ is a trainable time-step size parameter. Since Eq. (\ref{eq:discrete-SSM}) becomes a recurrence in the hidden state vector $x_k$, LSSL can now be considered a special RNN with linearity. By setting the initial state to be $x_{-1} = 0$ and unfolding Eq.~ref{eq:continuous-SSM}):
\vspace{-1.0mm}
\begin{equation}
        v_k  =\boldsymbol{C \overline{A}}^k \overline{\boldsymbol{B}} u_0+\boldsymbol{C \overline{A}}^{k-1} \overline{\boldsymbol{B}} u_1+\cdots+\boldsymbol{C \overline{B}} u_k + \boldsymbol{D}u_k
\end{equation}
That is, the output sequence $\boldsymbol{v} = (v_0, v_1,..., v_{L-1})$ with length $L$ could be computed as a convolution $\boldsymbol{v} =  \boldsymbol{\overline{K}} * \boldsymbol{u} + \boldsymbol{D} \boldsymbol{u}$, where  $\boldsymbol{u} = (u_0, u_1,..., u_{L-1})$ and $\boldsymbol{\overline{K}}$ is called an SSM convolution kernel defined as a 1-D sequence:
%\begin{equation}
  %  \begin{aligned}
    %    \overline{\boldsymbol{K}} &:= (\overline{K}_0, \overline{K}_1,...,\overline{K}_{L-1})\\
      %  &=\left(\boldsymbol{C \overline{B}}, \boldsymbol{C \overline{A B}}, \ldots, \boldsymbol{C \overline{A}}^{L-1} \overline{\boldsymbol{B}}\right)
   % \end{aligned}
%\end{equation}
\begin{equation}
        \overline{\boldsymbol{K}} := (\overline{K}_0, \overline{K}_1,...,\overline{K}_{L-1})=\left(\boldsymbol{C \overline{B}}, \boldsymbol{C \overline{A B}}, \ldots, \boldsymbol{C \overline{A}}^{L-1} \overline{\boldsymbol{B}}\right)
\end{equation}

Therefore, LSSL can also be viewed as a convolutional layer with a global receptive field~\cite{gu2021combining, Gu2022efficiently}, and it can be computed very efficiently with fast Fourier transform once the SSM convolutional kernel, $\overline{\boldsymbol{K}}$, is known.

One major bottleneck of LSSL is that treating $\boldsymbol{A}$ as trainable parameters means computing $\boldsymbol{\overline{K}}$ as many times as it is necessary in the training stage. However, it is challenging to compute $\overline{\boldsymbol{A}}^i$ efficiently enough for practical usage without making the whole model unstable. The S4 layer overcomes this critical issue by decomposing the state transition matrix $\boldsymbol{A}$ into a sum of a low-rank matrix~\cite{markovsky2008} and a skew-symmetric matrix~\cite{bunch1982}. In practice, S4 re-parameterizes the state-transition matrices $\mathbf{A}$ as $\boldsymbol{A} = \boldsymbol{\Lambda} - \boldsymbol{PP^*}$, where $\boldsymbol{\Lambda} \in \mathcal{C}^{N}$ is a diagnoal matrix and $\boldsymbol{P} \in \mathcal{C}^N$. In S4,  $\boldsymbol{D}$ is set equal to $\boldsymbol{0}$  since it could be replaced by a residual connection. Therefore, an S4 layer is comprised of 4N trainable matrix parameters: $\boldsymbol{\Lambda, P, B, }$ and $\boldsymbol{C}$. The interested reader is referred to~\cite{Gu2022efficiently, Gu2022Sashimi} for more details on S4.

\subsection{SSM with Multi-dimensional Patterns}
\label{subsec:multi-dim S4}
The S4 layer was developed for 1-D inputs, which limited its applicability. In ~\cite{gu2021combining, Gu2022efficiently, Gu2022Sashimi}, that limitation was overcome by (i) running $H$ independent copies of the S4 layer on a 2-D input features with a shape of $(H, L)$, and (ii) mixing all output features by a position-wise feedforward layer. In practice, the 2-D input is assumed to consist of $H$ independent signals, analogous to a 1-D CNN layer with H channels. However, most of the 2-D inputs are correlated with each other in both axes. For example, a 2-D spectrogram is not only time-dependent but also has strong spectral dependencies across the frequency axis. In \cite{Nguyen2022snd}, the conventional S4 layer was extended to multi-dimensional signals by turning the standard SSM (1-D ODEs) into multi-dimensional partial differential equations (PDEs) governed by an independent SSM in each dimension. To make it clear, let $u = u(t_1, t_2) \in \mathcal{R}^2 \rightarrow \mathcal{C}$ and $v = v(t_1, t_2) \in \mathcal{R}^2 \rightarrow \mathcal{C}$ be the input and output signals, and $x(t_1, t_2)$  be the SSM state with dimension $N_1 \times N_2$, which equals to the outer product of $x_1(t_1, t_2) \in \mathcal{C}^{N_1}$ and $x_2(t_1, t_2) \in \mathcal{C}^{N_2}$. A 2-D SSM thus can be represented by the following two-variable PDEs:
\vspace{-2mm}
\begin{equation}
\label{eq:S4ND-PDEs}
    \begin{aligned}
    \frac{\partial}{\partial t_1} x(t_1, t_2 ) &=\boldsymbol{A}_1 x_1(t_1, t_2) \otimes x_2(t_1, t_2)+\boldsymbol{B}_1 u(t_1, t_2) \\
    \frac{\partial}{\partial t_2} x(t_1, t_2) & =x_1(t_1, t_2) \otimes \boldsymbol{A}_2 x_2\left(t_1, t_2\right)+\boldsymbol{B}_2 u\left(t_1, t_2\right) \\
    v(t_1, t_2) & =\langle\boldsymbol{C}, x(t_1, t_2)\rangle
    \end{aligned}
\end{equation}
where $\otimes$ is the outer-product operator, $\langle\cdot, \cdot\rangle$ is the point-wise inner-product operators for two matrices.
Factoring that matrix $\boldsymbol{C}$ as a low-rank tensor and using the standard 1-D S4 layer as a black box, a 2D-version S4 layer is equivalent to a multi-dimensional convolution with an infinite receptive field. The multi-dimensional S4 layer is referred to as S4ND~\cite{Nguyen2022snd}. The same bilinear method mentioned in Sec.~\ref{subsec:1DSSM_intro} can be applied when handling 2D discrete-time inputs.

\section{Proposed Deep Structured State Space Modeling for Speech Enhancement}

\subsection{Time-domain S4-based Model}
\label{subsec:time-domain S4 model}
The original S4-based model in~\cite{Gu2022Sashimi} can be used as a time-domain regression SE model,  consisting of  repeated S4-block combined with a U-net architecture.  This model is referred to as \textit{Time-domain S4 U-Net}. A 1-D convolutional layer with kernel size $=1$ will first turn the single-channel noisy speech into a hidden feature with 64 channels before it goes through the whole U-net path. Another single-kernel 1-D convolutional layer will instead shrink the channel size back to 1, which represents the enhanced speech. More details can be found~\cite{Gu2022Sashimi}.

The time-domain S4 U-Net is optimized by the loss function proposed in~\cite{Dfossez2020}, which is the L1 loss over the waveform together with a multi-resolution short-time Fourier transform (STFT) loss over the spectrogram magnitude~\cite{Yamamoto2020}:
\vspace{-1mm}
\begin{gather} 
    \operatorname{loss}_{S T F T}(y, \tilde{y})=\frac{\mid|S T F T(y)|-|S T F T(\tilde{y})|\mid_F}{\mid|S T F T(y)|\mid_F} \notag \\
    +\frac{1}{T}\mid\log (|S T F T(y)|)-\log (|S T F T(\tilde{y})|)\mid_1 \\
    \operatorname{loss}(y, \tilde{y})=\frac{1}{T}\mid y-\tilde{y}\mid_1+\frac{1}{M} \sum_{i=1}^M \operatorname{loss}_{S T F T}^{(i)}(y, \tilde{y})
\end{gather}
where $\operatorname{STFT}$ stands for short-time Fourier transform, $|\cdot|_{F}$ is the Forbenious norm, $y, \tilde{y}$ are the clean and estimated speech, respectively, and $M$ means the total number of different STFT settings when computing $\operatorname{loss}_{S T F T}(y, \tilde{y})$. We use the same three STFT settings to compute $\operatorname{loss}_{S T F T}^{(i)}(y, \tilde{y})$ in~\cite{Dfossez2020}.

\subsection{TF-domain S4-based Model}
\label{subsec:TF-domain S4 Model}
We deploy two S4-based SE models in the TF-domain, namely (i) only the magnitude of the spectrogram is enhanced and then the noisy phase is used to reconstruct the speech waveform, or (ii) the complex noisy spectrogram is used to directly estimate both the magnitude and phase information simultaneously. In the magnitude-only scenario, two variants are put forth: a \textit{regression} model, which maps noisy spectrograms to clean spectrograms, and a \textit{masking} model which predicts a spectrogram mask. In the complex scenario, we adopt the masking method for the complex spectrogram proposed in \cite{choi2019, hu2020} since the regression-based SE model led to an unstable training process in the complex case. In sum, there are three training scenarios: \textit{mag-regression}, \textit{mag-masking}, and \textit{complex-masking}.

L1 loss is employed to optimize our TF-domain deep architecture, as suggested in~\cite{Qi2020}. More specifically, the following loss is used when only the magnitude information is taken into account (i.e., \textit{mag-regression} and \textit{mag-masking}):
\begin{equation}
    \operatorname{Loss}_{mag} = \frac{1}{TF} \sum_{t=0}^{T-1} \sum_{f=0}^{F-1} \mid S(t, f) -\Tilde{S}(t, f)\mid
\end{equation}
where $S$ and $\Tilde{S}$ are the magnitude spectrograms of the clean speech and enhanced speech, respectively, and $T$ and $F$ are the numbers of frames and frequency bins, respectively.

When the complex spectrogram is used as the input/output, the loss function becomes:
\begin{equation}
    \begin{aligned}
        \operatorname{Loss}_{complex}= & \frac{1}{T F} \sum_{t=0}^{T-1} \sum_{f=0}^{F-1}
        \mid S_r(t, f)-\tilde{S}_r(t, f) \mid \\
        & +\mid
            S_i(t, f)- \tilde{S}_i(t, f)
        \mid + \operatorname{Loss}_{mag}
    \end{aligned}
\end{equation}
where $r$, and $i$ denote the real and imaginary parts of the complex-valued spectrogram, respectively.

\subsubsection{1-D S4-based Model with Whitening Transform}
As discussed in Section~\ref{subsec:multi-dim S4}, we could directly adopt the 1-D S4-based layer to build a deep SE model working on 2-D spectrograms by treating frequency bins as a group of independent 1-D signals. In doing so, the deep SE model can not exploit dependencies across the frequency axis. However, we introduced a whitening transformation~\cite{Kessy2015} to the input spectrogram so that the frequency bins can be considered statistically uncorrelated to one another. %Leveraging the whitening transformation will make the usage of a 1-D S4 layer more meaningful with respect to using the original spectrogram directly.
This system is called  \textit{TF-domain S4 U-Net}.

\begin{figure}[t]
    \centering
    \includegraphics[width=0.46\textwidth]{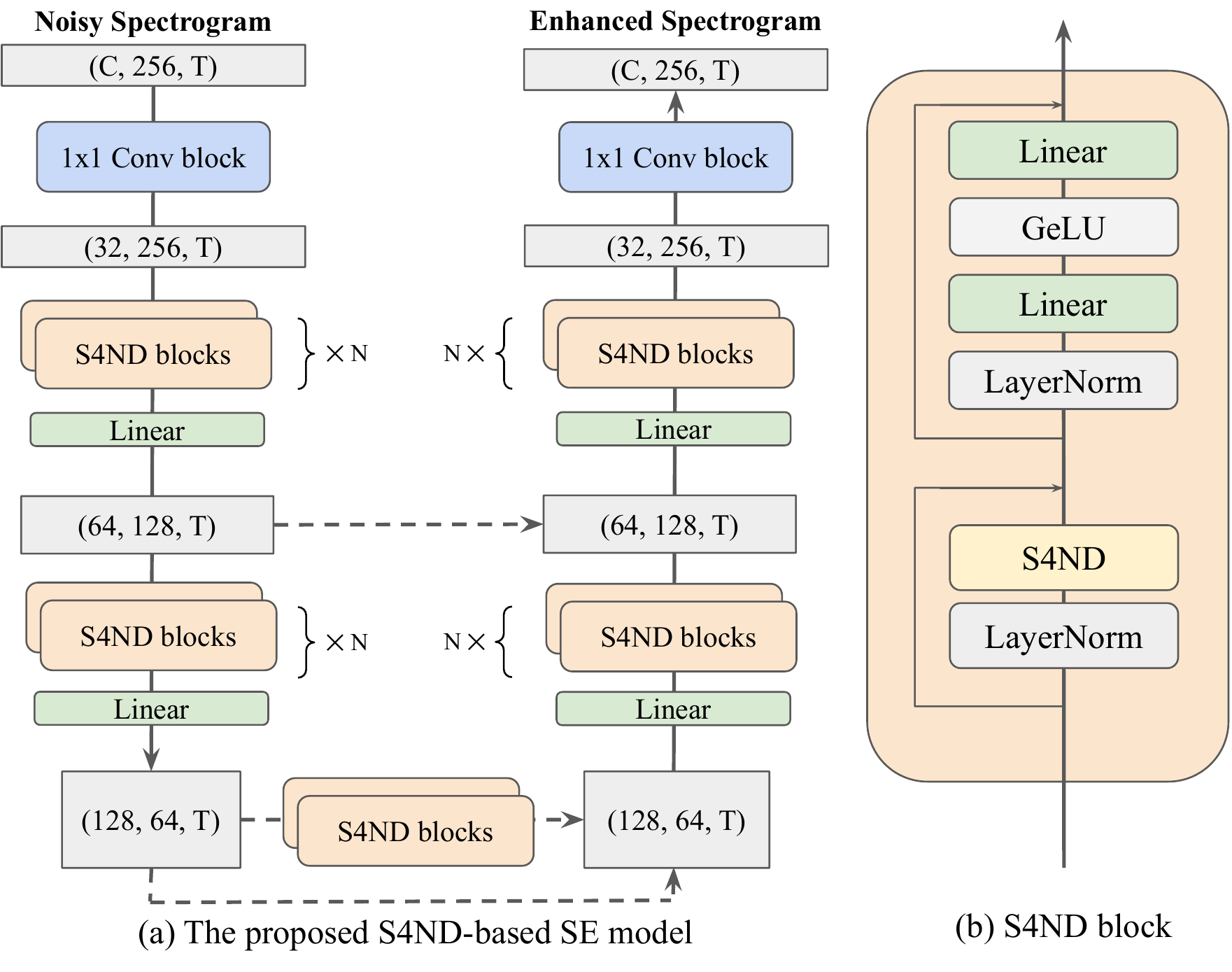}
    \caption{(a) The proposed S4ND U-Net. $C$  is either 1 or 2, depending on whether the input is a magnitude spectrogram ($C=1$) or a complex spectrogram ($C=2$), and $N$ is the number of the S4 blocks stacked per down/up layer. N=4 is used in all experiments. (b) The detailed structure of an S4ND block. The original S4 block's design~\cite{Gu2022Sashimi} was used but S4 was replaced by S4ND.}
    \label{fig:S4SE}
    \vspace{-3mm}
\end{figure}

\subsubsection{Multi-dimensional S4-based Model}
\label{subsubsec:S4ND-U-Net}
Although the whitening transformation can theoretically alleviate the frequency-dependent problem, it doesn't really help the model capture the 2-D information as well as a 2-D CNN layer does. Our experiments (see Sec.~\ref{subsec:results}) show that while adding the whitening transformation is helpful, it doesn't perform better than the time-domain S4-based model but with more model parameters. For that reason, we build a new SE model based on the S4ND layer, which is named \textit{S4ND U-Net}.

Fig.~\ref{fig:S4SE} illustrates how the proposed \textit{S4ND U-Net} is built. \textit{S4ND U-Net} is similar to the Deep Complex Convolutional Recurrent Network (DCCRN) model~\cite{hu2020}, which is a U-net-like architecture with several down-layers and up-layers, where we have however replaced all complex CNN/RNN layers with S4ND layers. Furthermore, we have reduced the model complexity by shrinking the number of down-layers and up-layers to 2 (the original DCCRN model had 6 layers), since it is unnecessary to stack too many S4ND layers to increase the receptive field. With complex spectrograms, we stack the real and imaginary parts as a real-value spectrogram with two channels.

\section{Experiments and Result Analyses}
\subsection{Experimental Setup}
\subsubsection{Dataset}
We choose the VoiceBank-DEMAND dataset~\cite{ValentiniBotinhao2017}, a common SE benchmark data set providing recordings from 30 speakers with 10 types of noise, to assess our models. The data set is split into a training and a testing set with 28 and 2 speakers, respectively. Four types of signal-to-noise ratios (SNRs) are used to mix clean samples with noise samples in the dataset, [0, 5, 10, 15] dB for training and [2.5, 7.5, 12.5, 17.5] dB for testing. We follow the approach used by Lu et al.~\cite{Lu2022} to form the validation set by excerpting two speakers from the training set. This resulted in 10,802 utterances for training and 770 for validation. The testing set included a total of 824 utterances. All recordings were downsampled from 48 kHz to 16 kHz. Our evaluation metrics include the wide-band perceptual evaluation of speech quality (PESQ)~\cite{Rix2001}, prediction of the signal distortion (CSIG), prediction of the background intrusiveness (CBAK), and prediction of the overall speech quality (COVL).

\subsubsection{STFT feature transformation and data augmentation}
\label{subsec:data-aug}
During training, an amplitude transformation as in ~\cite{richter2022} is used to normalize the amplitude of all complex coefficients of both noisy and clean spectrograms. Two out of four data augmentation methods in \cite{Dfossez2020}, namely, \textit{Remix} and \textit{BandMask} are used. \textit{Remix} shuffles the noises within one batch to form new noisy mixtures. \textit{BandMask} is a band-stop filter that removes 20\% of the frequencies starting from $f_0$, a random frequency sampled uniformly in the mel scale.

\subsection{Experimental Results}
\label{subsec:results}
In Table~\ref{tab:Time-domain-results}, we first assess our baseline \textit{time-domain S4 U-Net} discussed in Sec.~\ref{subsec:time-domain S4 model} and compare it with other time-domain techniques. We build \textit{time-domain S4 U-Net} according to the default hyper-parameter setting in~\cite{Gu2022efficiently}. A visual inspection of Table~\ref{tab:Time-domain-results} shows that when compared with other models, time-domain S4 U-Net achieves a good PESQ (with PESQ=3.02 in the bottom row) and the best CSIG, CBAK, and COVL scores. Although DEMUCS (with PESQ=3.07 in the fourth row) has a better PESQ score than \textit{time-domain S4 U-Net}, its performance degrades to 2.93 when its model size shrinks from 33.5 million to 18.9 million. This result verifies that the S4 layers can replace CNN/RNN layers in a DNN-based model to reduce the model sizes while maintaining competitive performances.

\begin{table}[t]\footnotesize
    \centering
        \caption{Evaluation results of different time-domain U-Net models. Note that we utilize only parts of the data augmentation methods in DEMUCS~\cite{Dfossez2020}.}
        \vspace{-2mm}
    \label{tab:Time-domain-results}

    \begin{adjustbox}{width=\columnwidth}
        \begin{tabular}{c|c|cccc}
            \toprule
            Model             & \begin{tabular}[c]{@{}c@{}}Params\\ (Million)\end{tabular} & PESQ & CSIG & CBAK & COVL \\ \midrule
            Noisy             & n/a                                                        & 1.97 & 3.35 & 2.44 & 2.63 \\ \midrule \midrule
            Wave-U-Net~\cite{Macartney2018}        & 10.0                                  & 2.40 & 3.52 & 3.24 & 2.96 \\
            Attention Wave-U-Net~\cite{Giri2019}   & -                                     & 2.62 & 3.91 & 3.35 & 3.27 \\
            DEMUCS (small)~\cite{Dfossez2020}      & 18.9                                  & 2.93 & 4.22 & 3.25 & 3.52 \\
            DEMUCS (large)~\cite{Dfossez2020}      & 33.5                                  & 3.07 & 4.31 & 3.4  & 3.63 \\ \midrule \midrule
            Time-domain S4 U-Net & \multirow{2}{*}{7.14}                                   & 2.97 & 4.36 & 3.49 & 3.65 \\
            + data augmentation  &                                                         & 3.02 & 4.45 & 3.56 & 3.75 \\ \bottomrule
        \end{tabular}
   \end{adjustbox}
   \vspace{-3mm}
\end{table}

\begin{table}[t]\footnotesize
    \centering
        \caption{A comparison of different TF-domain models.}
        \vspace{-2mm}
    \label{tab:TF-domain-results}
   \begin{adjustbox}{width=\columnwidth}
        \begin{tabular}{c|c|cccc}
            \toprule
            Model                                  & \begin{tabular}[c]{@{}c@{}}Params\\ (Million)\end{tabular} & PESQ & CSIG & CBAK & COVL \\ \midrule
            Noisy                                  & n/a                                                        & 1.97 & 3.35 & 2.44 & 2.63 \\ \midrule \midrule
            DCCRN~\cite{hu2020, Lv2022}            & 3.7                                                        & 2.54 & 3.74 & 3.13 & 2.75 \\
            %DCCRN+~\cite{lv2021}                   & 3.3                                                        & 2.84 & -    & -    & -    \\
            S-DCCRN~\cite{Lv2022}                  & 2.34                                                       & 2.84 & 4.03 & 3.43 & 2.97 \\
            DCUnet-10~\cite{choi2019}              & 1.4                                                        & 2.72 & 3.74 & 3.60 & 3.22 \\
            DCUnet-16~\cite{choi2019}              & 2.3                                                        & 2.93 & 4.10 & 3.77 & 3.52 \\
            DCUnet-20~\cite{choi2019}              & 3.5                                                        & 3.13 & 4.24 & \textbf{4.00} & 3.69 \\
            Metric GAN+~\cite{Fu2021}              & 2.6                                                        & \textbf{3.15} & 4.14 & 3.16 & 3.64 \\ \midrule \midrule
            TF-domain S4 U-Net                     & \multirow{3}{*}{17.85}                                     & 2.96 & 4.24 & 3.49 & 3.60 \\
            + data augmentation                    &                                                            & 3.05 & 4.34 & 3.54 & 3.69 \\
            + whitening                            &                                                            & 3.07 & 4.35 & 3.56 & 3.72 \\ \midrule
            S4ND U-Net                             & \multirow{2}{*}{\textbf{0.75}}                             & 2.99 & 4.37 &  3.53 & 3.70  \\
            + data augmentation                    &                                                            & \textbf{3.15} & \textbf{4.52} & 3.62 & \textbf{3.85} \\
            \bottomrule
        \end{tabular}
    \end{adjustbox}
    \vspace{-4mm}
\end{table}

Next in Table~\ref{tab:TF-domain-results}, we compare the proposed \textit{TF-domain S4 U-Net} and \textit{S4ND U-Net} with previously proposed TF-domain models. All networks use a U-Net structure and the same complex-masking approach~\cite{choi2019} except Metric GAN+. %By comparing the results in Tables~\ref{tab:Time-domain-results} and~\ref{tab:TF-domain-results}, we can see that time-domanin S4 U-Net performs much better than TF-domain regardless of the whitening transformation, as expected.
From Table~\ref{tab:TF-domain-results}, we can see that S4ND U-Net attains the best results among all models on PESQ/CSIG/COVL metrics with relatively good CBAK scores. It is worth noting that Metric GAN+ optimizes the PESQ score directly (with the best PESQ=3.15 in the sixth row) at the cost of getting worse results on the other metrics. Additionally, DCUnet-20 attains the best CBAK of 4.00 with PESQ=3.13 in the fifth row. However, its evaluation results significantly drop when halving the number of layers. In contrast, S4ND U-Net uses only 0.75M parameters, corresponding to a 78.6\% reduction in model size compared to DCUnet-20. It also achieves a PESQ score of 3.15 (equal to the best GAN+) as shown in the bottom row.

To have a fair comparison, we retrain DCCRN from scratch with the same data augmentation and amplitude transformation (see Sec.~\ref{subsec:data-aug}) used to build S4ND U-Net. Two STFT settings (i.e., subscripts "$_1$" and "$_2$" in Table~\ref{tab:DCCRN-and-S4ND-U-Net}) are adopted to investigate how DCCRN and S4ND U-Net react when the input STFT spectrograms have different parameter settings. Furthermore, DCCRN parameters were limited to 1.21M by halving the number of down/up layers from 6 to 3. As shown in Table~\ref{tab:DCCRN-and-S4ND-U-Net}, S4ND U-Net outperforms both the small and large DCCRN under the first STFT setting, with PESQ scores of 3.15 and 3.18 in the fifth and sixth rows, respectively. When a bigger hop-length=255 is chosen and the time-length of input spectrograms decreases correspondingly (i.e., the second STFT setting), the S4ND U-Net's performance drops a little bit (with PESQ=3.09 and 3.11 in the bottom two rows) but remains superior to DCCRN. This suggests our S4ND U-Net works particularly well when dealing with long input sequences. With a large S4ND U-Net with 4.4M parameters by increasing the number of down/up layers from 2 to 5, We only improve PESQ slightly from 3.15 to 3.18. This confirms there is no need to stack too many S4ND layers to increase the receptive field. Even without data augmentation, S4ND U-Net (with PESQ=2.99 shown in the second last row of Table~\ref{tab:TF-domain-results}) still outperforms both DCCRN and DCUnet-16.

\begin{table}[t]\footnotesize
    \caption{Evaluation results of models with two sizes (S \& M) and two STFT settings (subscripts "$_1$" \& "$_2$"). "$_1$" means we adopt 510/400/100 for n-fft/win-length/hop-length, while "$_2$" means 510/255/255 was adopted. Note that we set n-fft=510 to make the frequency dimension a ratio of two.}
    \vspace{-2mm}
    \label{tab:DCCRN-and-S4ND-U-Net}
    \centering
    \begin{adjustbox}{width=0.95\columnwidth}
    \begin{tabular}{c|c|cccc}
        \toprule
        \begin{tabular}[c]{@{}c@{}}Model (size) \end{tabular}               & \begin{tabular}[c]{@{}c@{}}Params\\ (Million)\end{tabular} & PESQ & CSIG & CBAK & COVL \\ \midrule
        DCCRN$_1$ (S)                                                                        & 1.19         & 2.75 & 3.62 & 2.99 & 3.17
        \\
        DCCRN$_1$ (M)                                                                & 3.48         & 2.87 & 3.93 & 2.59 & 3.39 \\ \midrule
        DCCRN$_2$ (S)                                                                        & 1.19         & 2.69 & 3.62 & 3.35 & 3.14 \\
        DCCRN$_2$ (M)                                                                & 3.48         & 2.83 & 3.91 & 2.51 & 3.36 \\ \midrule\midrule
        S4ND U-Net$_1$ (S)                                                                   & 0.75         & 3.15 & 4.52 & 3.62 & 3.85 \\
        S4ND U-Net$_1$ (M)                                                               & 4.44         & 3.18 & 4.49 & 3.63 & 3.85 \\ \midrule
        S4ND U-Net$_2$ (S)                                                                   & 0.75         & 3.09 & 4.44 & 3.60 & 3.78 \\
        S4ND U-Net$_2$ (M)                                                                & 4.44         & 3.11 & 4.46 & 3.59 & 3.79 \\ \bottomrule
    \end{tabular}
    \end{adjustbox}
    \vspace{-2mm}
\end{table}

\begin{table}[t]\footnotesize
    \centering
        \caption{A comparison of three TF-domain S4ND U-Nets.}
        \vspace{-2mm}
    \label{tab:S4ND with different scenarios}
  \begin{adjustbox}{width=0.9\columnwidth}
    \begin{tabular}{c|cccc}
        \toprule
        Scenario                    & PESQ & CSIG & CBAK & COVL \\ \midrule
        \textit{mag-regression}     & 3.05 & 4.44 & 3.47 & 3.76 \\
        \textit{mag-masking}        & 3.12 & 4.51 & 3.58 & 3.83 \\
        \textit{complex-masking}    & 3.15 & 4.52 & 3.62 & 3.85 \\ \bottomrule
    \end{tabular}
   \end{adjustbox}
   \vspace{-4mm}
\end{table}

%\subsubsection{Ablation Study}
Finally, we assess S4ND U-Net under three different TF-domain scenarios, namely \textit{mag-regression}, \textit{mag-masking}, and \textit{complex-masking}. Results in Table~\ref{tab:S4ND with different scenarios} show that \textit{complex-masking} attains a better PESQ score of 3.15 as shown in the bottom row when compared to 3.05 in the top and 3.12 in the middle rows. It is noted that although complex masking manages to achieve the best result among the three TF-domain scenarios compared here, models with only enhanced magnitudes plus noisy phase information for waveform reconstruction also perform well using the proposed S4ND U-net architecture.
%However, further investigation suggests that this improvement is not due to successful phase estimation or different model sizes. We identified this issue as a potential area for future research.

\section{Conclusion}
We have conducted a series of experiments to investigate new uses of S4 layers for speech enhancement. We first develop an S4-based deep SE neural model to enhance speech in the time domain and attain promising results. To extend the approach to multi-dimensional inputs, we propose two techniques: a whitening transformation and an S4ND U-Net architecture. We found that our proposed S4ND U-Net not only outperforms the original time-domain S4 U-Net but also attains comparable or better scores in four evaluation metrics when contrasted with other time-domain and TF-domain U-Net models. Our results also empirically verify that fewer S4ND layers can be used for compact model design, resulting in a robust TF-domain SE model with a limited size. Codes used in this work are released at ~\url{https://github.com/Kuray107/S4ND-U-Net_speech_enhancement}

\clearpage
\footnotesize
\bibliographystyle{IEEEtran}
\bibliography{mybib}

\end{document}